\documentclass[a4paper,12pt]{article}

\usepackage[dvips]{graphicx}

\begin{document}

\title{\large\bfseries
Constructing Effective Pair Wave Function 
from Relativistic Mean Field Theory with a Cutoff
}

\author{\large
Tomonori Tanigawa$\,^{1,}$\thanks{E-mail: tomo2scp@mbox.nc.kyushu-u.ac.jp}
~and Masayuki Matsuzaki$\,^{2,}$\thanks{E-mail: matsuza@fukuoka-edu.ac.jp}\\
{\small\itshape $^{1}$Department of Physics, Kyushu University, Fukuoka 812-8581}\\
{\small\itshape $^{2}$Department of Physics, Fukuoka University of
Education, Munakata 811-4192}
}

\date{}

\maketitle

\begin{abstract}
We propose a simple method to reproduce the $^1S_0$ pairing properties of 
nuclear matter, which are obtained using a sophisticated model, by introducing a 
density-independent cutoff into the relativistic mean field model. 
This can be applied successfully to the physically relevant density range.
\end{abstract}

 The $^1S_0$ pairing gap $\Delta$ in infinite nuclear matter is obtained by 
solving the gap equation, 
\begin{equation}
  \Delta(p)=-{1\over{8\pi^2}}
  \int_0^\infty \bar v(p,k)
         {\Delta(k)\over\sqrt{(E_k-E_{k_{\rm F}})^2+\Delta^2(k)}}k^2dk\, ,
\label{eq1}
\end{equation}
with $\bar v$ indicating the antisymmetrized matrix elements of the 
particle-particle interaction $v_{\rm pp}$. One can see from this equation 
that the physical ingredients are the single-particle energies $E_k$ and 
$v_{\rm pp}$. In sophisticated microscopic approaches, the $E_k$ are obtained from 
Brueckner-Hartree-Fock calculations with bare $N$-$N$ interactions, which are 
fitted to the phase shifts of the $N$-$N$ scatterings in free space. 
As for $v_{\rm pp}$, most calculations employ bare interactions, while 
some others employ medium-renormalized interactions, such as the \mbox{$G$ matrices}. 
Approaches involving calculations of the former type are based on the view that 
the gap equation itself possesses 
a mechanism to evade strong short-range repulsions, and, accordingly, use 
of medium-renormalized interactions results in a double 
counting~\cite{cor1,cor2,baldo,nrel4}. The forty-year history of 
non-relativistic studies of the pairing problem~\cite{nrel2}~has shown that 
all the bare $N$-$N$ interactions that are fitted to the phase shifts give 
almost identical pairing gaps for the $^1S_0$ channel. This is because 
a separable approximation~\cite{bj}~can be made for the \mbox{$S$ wave} channels 
in which a virtual ($^1S_0$) or real ($^3S_1$) bound-state pole exists in 
the \mbox{$T$ matrices}~\cite{carl}, and this leads to an approximate relation 
between the pairing gap and the phase shift applicable to the low-density 
region~\cite{shift}. Medium renormalizations are understood to cause the gap 
to become larger because they weaken the short-range repulsion. Irrespective 
of whether the medium renormalizations are included, the 
particle-hole polarizations should be considered in the next order 
according to a diagrammatic analysis of the gap equation~\cite{mig}, and 
it is said that they act to reduce the gaps~\cite{cbf,pol1,pol2}.

 Another approach to the pairing in nuclear matter is based on the effective 
interactions that are constructed from the beginning to describe finite-density 
systems. An example is the Gogny force~\cite{gogny}, which 
describes the bulk and the pairing properties of infinite matter quite well 
without any cutoffs~\cite{nrel1}, and another is represented by various versions 
of the Skyrme forces, which require cutoffs for the description of the 
pairing~\cite{koma}.
From the viewpoint of the double counting of the short-range correlation 
mentioned above, however, the adequateness of the use of effective forces 
in the particle-particle (p-p) channel is not evident~\cite{nrel4}.
Although this is still an open problem, the Gogny force is said to act 
as a bare force in the p-p channel~\cite{be}.

Similarly to the studies discussed above, the first relativistic study of the 
pairing in 
nuclear matter was carried out in 1991~\cite{kr}~by adopting a phenomenological 
interaction, the relativistic mean field (RMF) model, which succeeded in 
reproducing the bulk properties of the finite-density nuclear many-body 
systems. But the resulting pairing gaps were about three times larger than those 
accepted as standard in the non-relativistic studies. After a five-year blank, 
various attempts to improve this result have begun. These attempts can be 
classified into two groups: The first one employs $v_{\rm pp}$ 
which are consistent with the particle-hole (p-h) channel, 
i.e. the single-particle states~\cite{rel1,rel2,mr,mm}, and 
the second one employs $v_{\rm pp}$ which 
are not explicitly consistent with the p-h channel~\cite{rr,mt}.
In addition to these works which are based on the single-particle states 
of the RMF model, there exists another~\cite{rel3}~which is based on the 
single-particle states obtained through the Dirac-Brueckner-Hartree-Fock (DBHF) 
calculation~\cite{bonn}. We refer to this as the third type hereafter. 
The result that the calculations of the second and 
the third types give almost identical pairing gaps indicates that the pairing
properties are determined predominantly by the choice of the 
p-p channel interaction, irrespective of the details of the 
single-particle states. In addition, the feature that the obtained gaps are 
very similar to those given by the non-relativistic calculations adopting 
bare interactions in the p-p channel supports this 
further. As for the first type, a more elaborate 
calculation, such as one including the $N$-$\bar N$ polarizations, would be 
necessary~\cite{mr}. As a complement to this kind of study, however, simpler 
method suitable for realistic applications are also desirable. Examples for which 
realistic pairing strengths are indispensable are studies of the crust 
matter in neutron stars and finite open-shell nuclei. In particular, aside 
from the 
practical successes of the ``relativistic" Hartree-Fock-Bogoliubov (HFB) 
calculations implemented by a non-relativistic force~\cite{rhfb},
tractable relativistic $v_{\rm pp}$ derived from the Lagrangian of the 
RMF model are needed to keep the concept of the HFB calculation.

 The purpose of this paper is to construct a relativistic effective force 
which can be used also in the p-p channel as the Gogny force in the 
non-relativistic calculation. Therefore, first of all, we consider 
the difference between an effective force and a bare force. 
In Fig. 1(a) the one-boson exchange $v_{\rm pp}$ with the coupling constants 
of the $\sigma$-$\omega$ model, which is the simplest version of the RMF 
model, is shown in comparison with the Bonn-B potential~\cite{bonn},
which is an example of the relativistic bare $N$-$N$ interactions. 
Their shapes differ greatly. This is because the 
former is constructed so as to reproduce the saturation property without 
the short-range correlations, while the latter reproduces it in the DBHF 
calculation which implies them. This leads to the characteristic feature of the 
former that both the small-momentum negative {\it off-diagonal} matrix elements 
and the large-momentum positive ones are stronger than those for the latter. 
Both of them enhance the pairing gap, as discussed below. 

 The momentum integration in Eq. (\ref{eq1}) should run to infinity when bare 
$N$-$N$ interactions are adopted. In contrast, there is room to introduce 
a momentum cutoff when we adopt some phenomenological interactions, 
which are meaningful only for small momenta, as the Skyrme force. 
Evidently the assumption of the RMF model that the nucleon is a 
point particle cannot be justified at sufficiently large momenta. 
Combining this fact with the strong cutoff dependence in the 
momentum region 3 -- 8 fm$^{-1}$ in Fig. 1(b) suggests the possibility to 
choose a proper cutoff which 
describes the pairing gap quantitatively. Note that the necessity 
of cutting off the large-momentum repulsion in the $v_{\rm pp}$ derived 
from the RMF model has also been suggested in studies of medium-energy 
heavy-ion collisions~\cite{hi1,hi2}. This is interesting in the respect that 
two different phenomena, which involve large momentum transfers, suggest 
similar cutoffs in the RMF-based p-p interaction.

 In order to describe superfluidity quantitatively, not only the pair wave 
function 
\begin{equation}
\phi(k)={1\over2}{{\Delta(k)}\over \sqrt{(E_k-E_{k_{\rm F}})^2+\Delta^2(k)}}
\, , 
\label{eq2}
\end{equation}
which determines the gap at the Fermi surface
\begin{equation}
  \Delta(k_{\rm F})=-{1\over{4\pi^2}}
  \int_0^\infty \bar v(k_{\rm F},k)\phi(k)k^2dk\, ,
\label{eq3}
\end{equation}
but also its derivative, which determines the coherence length~\cite{coh} 
\begin{equation}
\xi=\left({{\int_0^\infty\vert{{d\phi}\over{dk}}\vert^2k^2dk}
            \over{\int_0^\infty\vert\phi\vert^2k^2dk}}\right)
      ^{1\over2}\, ,
\label{eq4}
\end{equation}
should be reproduced. The latter quantity measures the spatial size of the 
Cooper pair. In weakly-coupled systems, in which $\Delta(k_{\rm F})$ is 
determined by the diagonal matrix element $v(k_{\rm F},k_{\rm F})$ only and 
$\xi\gg d$ (where $d$ is the interparticle distance), $\Delta(k_{\rm F})$ 
and $\xi$ are intimately related to each other. But this does not hold for 
nuclear many-body systems, and the off-diagonal matrix elements 
$v(k_{\rm F},k)$ play important roles. Therefore here we attempt to find a 
density-independent cutoff $\Lambda_{\rm c}$ for the upper bound of the 
integrals in Eqs. (\ref{eq3}) and (\ref{eq4}) so as to reproduce, in a wide 
density range, 
$\Delta(k_{\rm F})$ and $\xi$ obtained by adopting the Bonn-B potential. 
In other words, we attempt to introduce an extra parameter into the 
$\sigma$-$\omega$ model to fit the pairing properties described by a 
sophisticated model without changing the bulk properties. 

The outline of the numerical calculations is as follows: We start from the 
$\sigma$-$\omega$ model with the no-sea approximation, as we confirmed in 
Ref.~\cite{mm} that the Dirac sea effects were negligible. The parameters 
used are $M=$ 939 MeV, $m_\sigma=$ 550 MeV, $m_\omega=$ 783 MeV, 
$g_\sigma^2=$ 91.64, and $g_\omega^2=$ 136.2~\cite{sw}. The calculations were 
done for symmetric nuclear matter ($\gamma=$ 4) and pure neutron matter 
($\gamma=$ 2). The pairing gap at each 
momentum is calculated by the gap equation (\ref{eq1}) with $\Lambda_{\rm c}$ 
and the effective mass equation 
\begin{equation}
  M^\ast=M-{{g_\sigma^2}\over{m_\sigma^2}}{\gamma\over{2\pi^2}}
  \int_0^{\Lambda_{\rm c}} 
  {M^\ast\over\sqrt{k^2+M^{\ast\,2}}}v_k^2k^2dk\, .
\label{eq5}
\end{equation}
Equations (\ref{eq1}) and (\ref{eq5}) couple to each 
other through 
\begin{eqnarray}
  v^2_k&=&{1\over2}\left(1-
   {{E_k-E_{k_{\rm F}}}\over\sqrt{(E_k-E_{k_{\rm F}})^2+\Delta^2(k)}}
                    \right)\, ,
  \nonumber\\
  E_k&=&\sqrt{k^2+M^{\ast\,2}}+g_\omega\langle\omega^0\rangle\, .
\label{eq6}
\end{eqnarray}
We search for the value of ${\Lambda_{\rm c}}$ that minimizes
\begin{equation}
\chi^2={1\over 2N}\sum_{k_{\rm F}}
\left\{\left({{\Delta(k_{\rm F})_{\rm RMF}-\Delta(k_{\rm F})_{\rm Bonn}}
           \over{\Delta(k_{\rm F})_{\rm Bonn}}}\right)^2 \right.
 \left. + \left({{\xi_{\rm RMF}-\xi_{\rm Bonn}}
           \over{\xi_{\rm Bonn}}}\right)^2
\right\}\, .
\label{eq7}
\end{equation}
Here we assume equal weights for 
$\Delta(k_{\rm F})$ and $\xi$. The single-particle states are determined by 
the $\sigma$-$\omega$ model in both the ``RMF" and the ``Bonn" cases, as in 
Refs.~\cite{rr} and \cite{mt}. The summation with respect to $k_{\rm F}$ is 
taken as $k_{\rm F}=$~0.2, 0.3, $\cdot\cdot\cdot$, 1.2 fm$^{-1}$, 
i.e. $N=$ 11, since 
we do not anticipate that the present method is applicable to 
the $k_{\rm F}\sim$ 0 case, as discussed later. 

 We found  that ${\Lambda_{\rm c}}=$ 3.60 fm$^{-1}$ minimizes $\chi^2$ for 
$\gamma=$ 4. This value indicates that not only the small-momentum part, where 
$v(k_{\rm F},k)<$ 0 and $\Delta(k)>$ 0, but also the large-momentum part, where 
$v(k_{\rm F},k)>$ 0 and $\Delta(k)<$ 0, contribute (see Fig. 1(a)) as 
pointed out in Refs.~\cite{rr} and \cite{mt}. The cutoff smaller than 2 fm$^{-1}$ 
determined in Ref.~\cite{rel2} leads to cutting the ``repulsive" part 
completely, and this corresponds to choosing the plateau around 2 fm$^{-1}$ in 
Fig. 1(b), as proposed in Ref.~\cite{koma} in the case of the Skyrme force. 
The present result does not agree with these previous ones. Reference~\cite{rel1} 
reports a result different from ours. 
As discussed in Ref.~\cite{mm}, there are two reasons for this difference. 
One reason is that they adopted coupling constants which reproduced the 
saturation in the Hartree-Fock approximation, not the Hartree (so-called MFT) 
approximation. This leads to larger pairing gaps~\cite{bvc}. The other reason is 
the difference in the evaluation of the Dirac sea effects.

 Figures 2(a) and (b) show how well the $\sigma$-$\omega$ model with 
${\Lambda_{\rm c}}$ chosen above reproduces $\Delta(k_{\rm F})$ and $\xi$, 
respectively, obtained with the Bonn-B potential for symmetric nuclear matter. 
One can see some deviations between the two models both near 
$k_{\rm F}\sim$ 0.2 fm$^{-1}$ and $k_{\rm F}\sim$ 1.2 fm$^{-1}$. As for the 
former, it is quite reasonable that the present model based on the mean field 
picture for finite-density systems does not give a good fit. 
 Actually, in such an extremely dilute system, the effective-range 
approximation for the free scattering is quite good~\cite{shift}. As for the 
latter, the deviation results from the fact that the superfluid phase in the 
RMF model, as well as in the Gogny force case~\cite{rr}, disappears in a nearly 
$\Lambda_{\rm c}$-independent manner at somewhat larger $k_{\rm F}$ than in the 
Bonn potential case. This at the same time causes the overall peak shift of 
$\Delta(k_{\rm F})$ to larger $k_{\rm F}$ and makes $\xi$ at large $k_{\rm F}$ 
small. We should note, however, that the critical density or $k_{\rm F}$, where 
the pairing gap disappears, has not been fully discussed yet. 
The result for pure neutron matter is very similar, except that 
$\Delta(k_{\rm F})$ is somewhat larger as seen in Fig. 1(b), 
and the superfluid phase survives up to somewhat larger $k_{\rm F}$, due to 
larger values of $M^\ast$ than in the symmetric matter case. Consequently, the 
present method gives a good fit still for $k_{\rm F}\sim$ 1.2 fm$^{-1}$.
The density of neutron matter in the inner crust of neutron stars corresponds 
to 0.2 fm$^{-1}$ $\,\lower .6ex\hbox{$\buildrel<\over\sim$}\,$ $k_{\rm F}$ 
$\,\lower .6ex\hbox{$\buildrel<\over\sim$}\,$
1.3 fm$^{-1}$~\cite{nrel2}. Therefore the present simple method covers the 
greater part of this range. In finite nuclei, pairing occurs near the nuclear 
surface, where the density is lower than the saturation point.
The present method gives a good description of this region.

 To summarize, we proposed a method to reproduce the $^1S_0$ pairing 
properties of 
infinite nuclear matter, obtained using a sophisticated DBHF plus a full-range 
gap equation adopting the Bonn potential, by introducing a 
momentum cutoff into the gap equation with the relativistic mean field model. 
This method was shown to be applicable in a wide and physically relevant 
density range. 
This points the way to consistent (i.e., using the same 
interaction in the p-h and the p-p channels)
relativistic HFB studies of neutron stars and finite nuclei.
Finally, we remark that the cutoff we introduced is density-independent, 
since our approach is based on the density-independent --- except the 
dependence through $M^\ast$ --- RMF interaction 
which was determined at the saturation point. This deserves further 
investigation.

\section*{Acknowledgements}
We would like to thank Professor R. Tamagaki for useful comments.

\begin{figure}
\begin{center}
\Large FIGURES 
\end{center}
\vspace*{1cm}
\includegraphics[width=6.6cm,keepaspectratio]{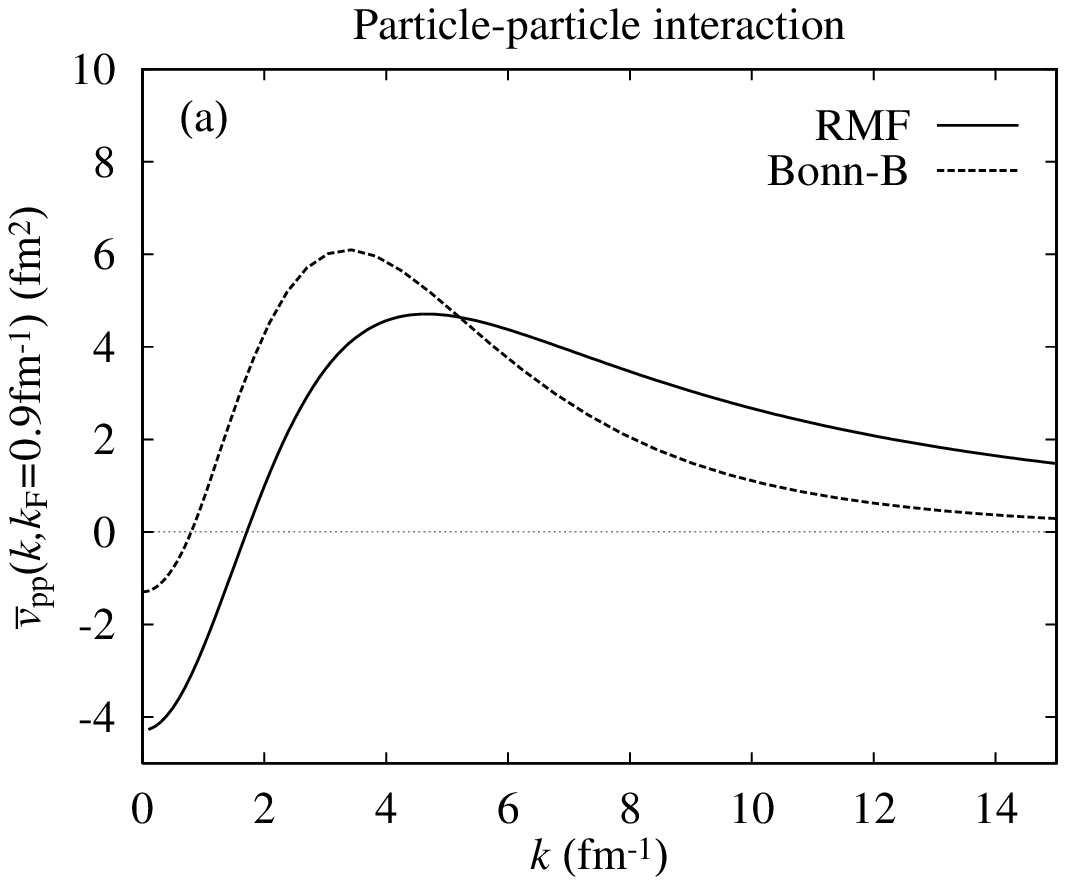}
\includegraphics[width=6.6cm,keepaspectratio]{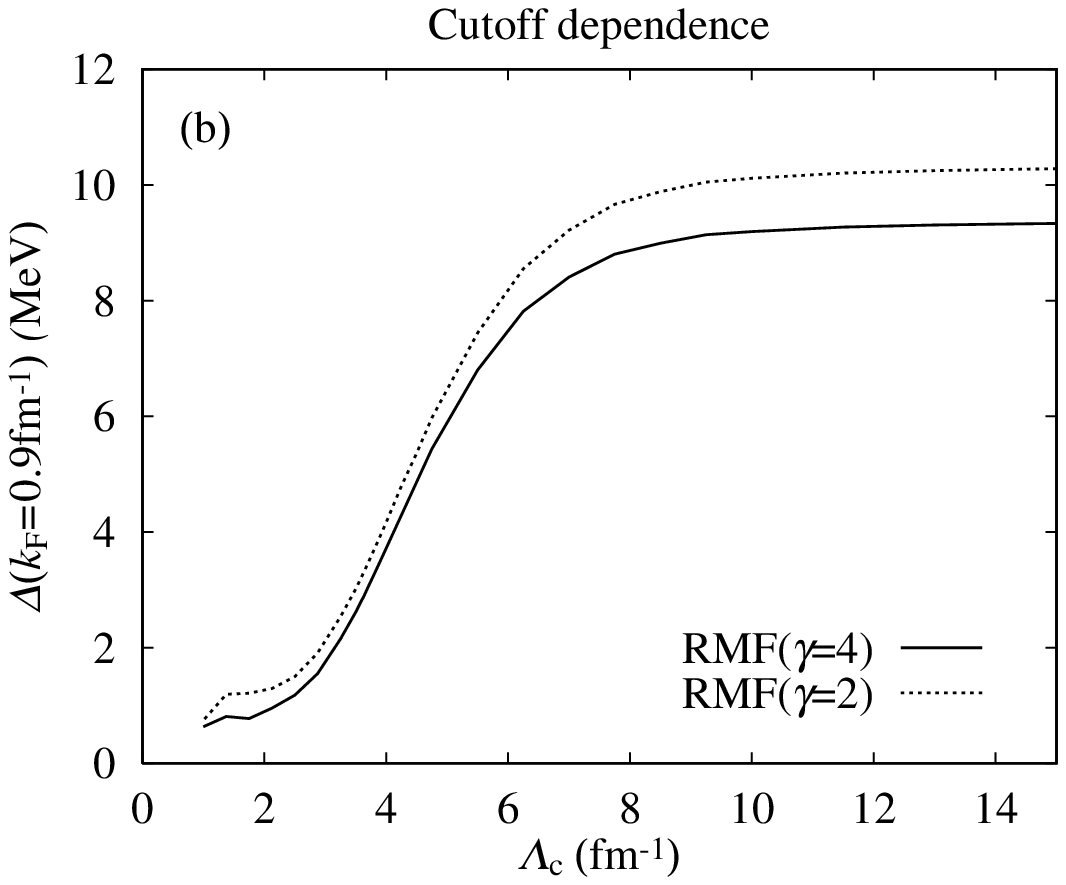}
\caption{
(a) Matrix element $\bar v_{\rm pp}(k,k_{\rm F})$ as functions of the 
momentum $k$, with the Fermi momentum $k_{\rm F}=$ 0.9 fm$^{-1}$. The solid and 
dashed curves indicate the results obtained by the relativistic mean field 
model and the Bonn-B potential, respectively. (b) Pairing gap at the Fermi 
surface, $k_{\rm F}=$ 0.9 fm$^{-1}$, obtained from the relativistic mean field 
model, as functions of the cutoff parameter in the numerical integrations. 
The solid and dotted curves indicate the results for symmetric nuclear matter and 
pure neutron matter, respectively. 
}
\vspace*{2cm}
\includegraphics[width=6.6cm,keepaspectratio]{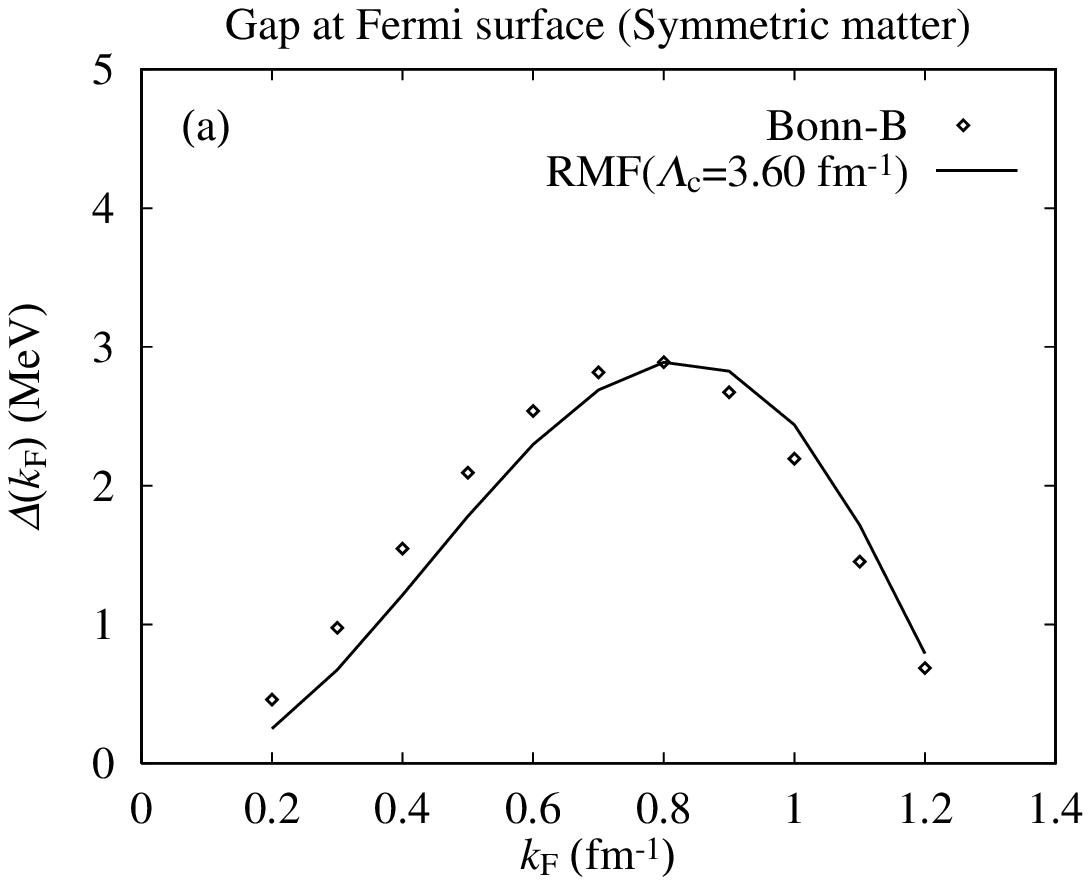}
\includegraphics[width=6.6cm,keepaspectratio]{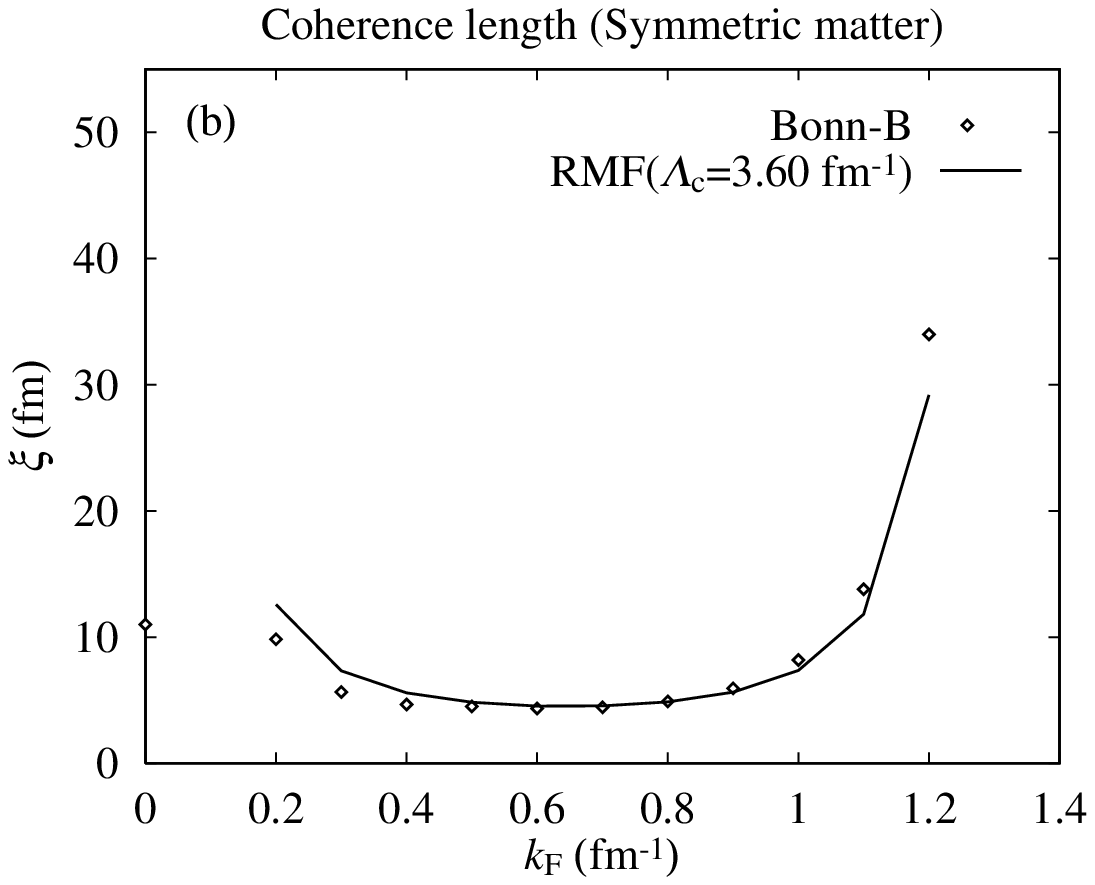}
\caption{
Pairing gap at the Fermi surface (a), and coherence length (b) as functions of 
the Fermi momentum, calculated for symmetric nuclear matter. The solid curves and 
diamonds indicate the results obtained using the relativistic mean field model 
with a momentum cutoff ${\Lambda_{\rm c}}=$ 3.60 fm$^{-1}$ and the Bonn-B 
potential, respectively.
}
\end{figure}

\end{document}